\documentclass[sigconf]{acmart}

\AtBeginDocument{%
  }

\setcopyright{acmlicensed}
\copyrightyear{2018}
\acmYear{2018}
\acmDOI{XXXXXXX.XXXXXXX}
\acmConference[Conference acronym 'XX]{Make sure to enter the correct
  conference title from your rights confirmation email}{June 03--05,
  2018}{Woodstock, NY}
\acmISBN{978-1-4503-XXXX-X/2018/06}

\usepackage{xcolor}
\usepackage{soul}    
\usepackage{hyperref}
\usepackage{todonotes}
\usepackage{booktabs}
\usepackage{multirow}
\usepackage{multicol}
\usepackage{subcaption}
\usepackage{caption}
\usepackage{enumitem}
\setlist[itemize]{leftmargin=1.5em}
\setlist[enumerate]{leftmargin=1.5em}
\sethlcolor{yellow}
\usepackage{tikz}

\newcounter{observation}
\newcommand{\observation}[1]{\refstepcounter{observation}
	\begin{center}
		\begin{tikzpicture}
			\node[
				draw=blue!40,
				fill=blue!10,
				text width=\linewidth,
				align=left
			] {#1};
		\end{tikzpicture}
	\end{center}
}

\begin{document}

\title{Does In-IDE Calibration of Large Language Models work at Scale?}

\author{Roham Koohestani}
\authornote{Both authors contributed equally to this research.}
\email{rkoohestani@tudelft.nl}
\affiliation{%
  \institution{Delft University of Technology}
  \city{Delft}
  \country{The Netherlands}
}

\author{Agnia Sergeyuk}
\authornotemark[1]
\email{agnia.sergeyuk@jetbrains.com}
\affiliation{%
  \institution{JetBrains Research}
  \city{Belgrade}
  \country{Serbia}
}

\author{David Gros}
\email{dgros@ucdavis.edu}
\affiliation{%
  \institution{University of California, Davis}
  \city{Davis}
  \state{California}
  \country{USA}
}

\author{Claudio Spiess}
\email{cvspiess@ucdavis.edu}
\affiliation{%
  \institution{University of California, Davis}
  \city{Davis}
  \state{California}
  \country{USA}
}

\author{Sergey Titov}
\email{sergey.titov@jetbrains.com}
\affiliation{%
  \institution{JetBrains Research}
  \city{Amsterdam}
  \country{The Netherlands}
}

\author{Prem Devanbu}
\email{ptdevanbu@ucdavis.edu}
\affiliation{%
  \institution{University of California, Davis}
  \city{Davis}
  \state{California}
  \country{USA}
}

\author{Maliheh Izadi}
\email{m.izadi@tudelft.nl}
\affiliation{%
  \institution{Delft University of Technology}
  \city{Delft}
  \country{The Netherlands}
}

\renewcommand{\shortauthors}{Koohestani et al.}

\begin{abstract}
The introduction of large language models into integrated development environments (IDEs) is revolutionizing software engineering, yet it poses challenges to the \textit{usefulness} and \textit{reliability} of Artificial Intelligence-generated code.
Post-hoc calibration of internal model confidences aims to align probabilities with an acceptability measure.
Prior work suggests calibration can improve alignment, but at-scale evidence is limited.
In this work, we investigate the feasibility of applying calibration of code models to an in-IDE context. 
We study two aspects of the problem:
(1) the \textit{technical} method for implementing confidence calibration and improving the reliability of code generation models, and
(2) the \textit{human-centered design} principles for effectively communicating reliability signal to developers.
First, we develop a \textbf{scalable} and \textbf{flexible} calibration framework which can be used to obtain calibration weights for open-source models using any dataset, and evaluate whether calibrators improve the alignment between model confidence and developer acceptance behavior. Through a large-scale analysis of over \textbf{24 million real-world} developer interactions across multiple programming languages, we find that a general, post-hoc calibration model based on Platt-scaling \textit{does not}, on average, improve the reliability of model confidence signals. We also find that while dynamically personalizing calibration to individual users \textit{can} be effective, its effectiveness is highly dependent on the \textit{volume} of user interaction data.
Second, we conduct a multi-phase design study with 3 expert designers and \textbf{153 professional developers}, combining scenario-based design, semi-structured interviews, and survey validation, revealing a clear preference for presenting reliability signals via non-numerical, color-coded indicators within the in-editor code generation workflow.
\end{abstract}



\maketitle

\section{Introduction}
\label{sec:Introduction}
The incorporation of Large Language Models (LLMs) into Integrated Development Environments (IDEs) has fundamentally altered the landscape of modern software development. Tools like GitHub Copilot~\cite{copilot2025tool}, JetBrains Junie~\cite{jetbrains2025junie}, and others have become increasingly sophisticated assistants, capable of generating everything from single lines of code to entire codebases and developer workflows~\cite{stackoverflow2025survey}. While these tools have increased developers' productivity~\cite{weisz2025examining,dohmke2023sea} and have thereby shortened development cycles, they also introduce several new concerns. These models, despite their impressive capabilities, are fallible and can generate unreliable code that is subtly incorrect, insecure, or inefficient~\cite{mohsin2024can,dou2024s}.

This introduces a critical challenge at the intersection of model capability and user interaction: how can a developer rely on and effectively collaborate with an Artificial Intelligence (AI) tool that is powerful yet imperfect? A developer's decision to accept, reject, or edit a suggestion depends on their trust in the model's output at any given moment. A crucial, yet often overlooked, aspect of this dynamic is \textit{model calibration}. In our work, \emph{model confidence} refers to the internal scores that a model assigns to its predictions (typically derived from token probabilities), while \emph{calibrated confidence} refers to a processed version of these probabilities that has been adjusted. A model is considered well-calibrated when its expressed confidence in a prediction directly corresponds to the actual likelihood of that prediction being correct
(in the sense that we explain below). Unfortunately, modern LLMs are notoriously poorly calibrated out-of-the-box, often exhibiting high confidence in erroneous generations. Research has demonstrated that LLMs consistently overestimate their correctness across diverse tasks~\cite{xiong2023llms,sun2025overconfident}. This miscalibration problem is particularly acute in code generation contexts, where studies show that generative code models are not well-calibrated without post-hoc adjustment~\cite{spiess2024calibration}.

This paper addresses the calibration of code LLM 
confidence, when used within an IDE, from both a technical and Human-AI user-experience perspective. We argue that calibration goes beyond being a technical challenge of statistical alignment, but is rather central to the usefulness and reliability of AI coding assistants. Accordingly, We evaluate the feasibility and utility of calibration from two connected perspectives in production IDE deployments: (1) the \textit{technical} method for implementing confidence calibration and improving the reliability of code generation models, and (2) the \textit{human-centered design} principles for effectively communicating reliability signals to developers. To this end, we aim to answer three main Research Questions (RQs).
\begin{itemize}
    \item \textbf{RQ 1.} Does calibrating confidence estimates better correlation
    (than raw confidence) with behavioral measures derived from telemetry in the wild?
    \item \textbf{RQ 2.} Does person-specific or project-specific model confidence calibration yield higher absolute correlation with behavioral measures?
    \item \textbf{RQ 3.} How do developers expect a reliability signal to be presented inside the IDE?
\end{itemize}

To address RQs 1 and 2, we conduct a wide-scale analysis of code completion system interactions inside the IDE. We analyze over \textbf{24 million} interaction records from more than \textbf{750,000 unique devices} across a \textbf{six-month} period. We additionally develop \textbf{Calibrate-CC}, a scalable and re-usable framework that can be used to obtain calibration weights for any open-source model using any dataset.

Our results show that post-hoc calibration generally \textit{does not} improve the reliability of model confidence scores (RQ1).
Furthermore, while personalized calibration for specific users or projects \textit{can} offer modest additional gains, we find that its effectiveness is highly dependent on the volume of interaction data, primarily benefiting the most active users (RQ2).

In our \textbf{multi-phase design user-study}, combining scenario-based design, semi-structured interviews, and larger-scale surveys with 3 expert designers and \textbf{153 professional developers} (RQ3), we find that developers have a clear preference for reliability to be communicated through non-numerical, color-coded indicators integrated directly into the in-editor code generation workflow.

In summary, the main contributions of our paper are:
\begin{itemize}
    \item a wide-scale in-the-wild evaluation of calibration frameworks for LLMs in the context of code completion.
    \item A scalable and flexible framework for model calibration based on commonly used libraries.
    \item A comprehensive user study on the needs and design of reliability indicators in a software development environment.
\end{itemize}

In the sections to follow, the background is presented in \autoref{sec:background}, followed by problem formulation in \autoref{sec:problem}. Research questions are addressed in \autoref{sec:rq1}, \autoref{sec:rq2}, and \autoref{sec:rq3}. We discuss implications and limitations in \autoref{sec:discussion} and conclude in \autoref{sec:conclusion}.

The replication package along with the code for CalibrateCC can be found at \href{https://zenodo.org/records/17433208}{https://zenodo.org/records/17433208}.

\section{Background}
\label{sec:background}

LLMs are increasingly used in IDEs to generate code suggestions and perform other software development tasks (\emph{e.g.}, Cursor~\cite{cursor2023}, JetBrains Junie~\cite{jetbrains2025junie}, Windsurf~\cite{windsurf2023}). While these tools promise productivity gains, their impact depends on the quality of suggestions that developers accept. Studies show that programmers working with AI assistants can unintentionally introduce insecure or incorrect code, sometimes with increased confidence in their outputs~\cite{perry2023users,gonzalez2025hilde}. This risk highlights the need for reliable signals that guide developers' acceptance and review decisions.

When using generative models for code completion, the model's confidence, computed as token probability, is often taken to be a proxy for likely correctness and is used when generating suggestions. However

in code generation, low probability may reflect multiple valid alternatives rather than incorrectness, and high probability can still coincide with errors.
\citet{vasconcelos2025generation} show that highlighting row low-probability tokens does not improve outcomes, while edit-based confidence models better align with developer needs. Other work similarly reports that raw model confidence is often miscalibrated, requiring improvement before it can serve as a useful guide~\cite{spiess2024calibration,kotti2025fools}.

Calibration methods aim to align predicted probabilities with observed correctness~\cite{spiess2024calibration}, and metrics such as expected calibration error and calibration curves are used to evaluate this alignment~\cite{steyvers2025metacognition}. Calibrated confidence scores provide developers with more reliable information for deciding when to accept or review suggestions. Complementary research also proposes lightweight intrinsic measures such as perplexity and entropy as early indicators of correctness or hallucination risk in generated code~\cite{kotti2025fools}.

Equally important is how reliability information is presented in the IDE. Human–AI interaction studies show that interface choices strongly influence trust and reliance~\cite{amershi2019guidelines,lee2004trust,mehrotra2024systematic}. Developers request confidence indicators that are granular but not overwhelming, such as highlights or compact visual cues~\cite{sun2022investigating,vasconcelos2025generation}. Explanations, sources, and consistency cues can further affect reliance: explanations often increase reliance on both correct and incorrect outputs, sources help reduce overreliance, and inconsistencies can serve as warnings~\cite{kim2025fostering}. Interactive approaches such as exposing local alternatives at points of model uncertainty have been shown to reduce vulnerabilities and give developers more control~\cite{gonzalez2025hilde}. Finally, confidence signals interact with human judgment. Recent work shows that developers' own self-confidence can align with AI confidence, sometimes persisting beyond the interaction~\cite{li2025confidence}. Without careful design, this alignment can lead to miscalibration on the human side. These findings emphasize that confidence is not only a model property but also a feature of the joint human–AI system in the IDE.

In summary, prior work shows that raw token probabilities are not
well-correlated with
test-passing correctness, that
calibrated confidence measures provide better 
signals, and that interface design shapes how such signals affect usage. What is missing are empirical studies that connect calibrated confidence to actual acceptance behavior in real-world IDE workflows and that compare alternative user interface designs for presenting these signals. Our study addresses this gap.

\section{Defining the Problem}
\label{sec:problem}

To analyze the calibration of code completion models, we must establish a formal framework that defines the core components of the developer-AI interaction. This framework provides the foundation for our research questions and subsequent analysis.

Let the state of the IDE at the moment a code completion is triggered be the \textbf{context}, denoted as $C$. A code generation model, $M$, takes $C$ and produces a \textbf{suggestion}, $S$, which is a sequence of tokens $(t_1, t_2, \ldots, t_k)$ intended to complete the developer's work.

\[S = M(C)\]

Internally, the model $M$ computes a score for the suggestion $S$, which we refer to as the raw \textbf{model confidence}, $\mathit{conf}(S, C)$. While this function can be implemented in various ways, in this work, we define it as the average log-probability of the generated tokens, which is equivalent to the geometric mean of their probabilities. This is formally expressed as:

\[
\mathit{conf}(S, C) = exp(\frac{1}{k} \sum_{i=1}^{k} \log P(t_i | C, t_1, \ldots, t_{i-1}))
\]

The goal is to assess whether the model's confidence score reflects the empirical reliability of its suggestion. We define an \textbf{outcome variable} $Y \in [0, 1]$, which quantifies the realized quality or correctness of a suggestion. Depending on the evaluation setting, $Y$ may be binary (e.g, accepted/rejected) or continuous (e.g., 
normalized similarity of suggested code to code as finally used)

A model achieves \textbf{perfect calibration} when, for each confidence value $p \in [0, 1]$ the expected result is equal to the reported confidence.
\[
E[Y \mid \mathit{conf}(S, C) = p] = p
\]

A model is said to be \textbf{miscalibrated} when this condition is not met, that is, when its expressed confidence is a poor estimator of the expected outcome. Modern LLMs are notoriously miscalibrated and often exhibit systematic overconfidence ($E[Y] < p$) or underconfidence ($E[Y] > p$); this was found
to be the case when $Y$
indicates ``test-passing
correctness''~\cite{spiess2024calibration}. 

One of our goals here is to learn \textbf{calibration model} (or calibration function), denoted as $f$. This model 
transforms the raw, uncalibrated model confidence into a new, calibrated probability, $\hat{p}$.
\[
\hat{p} = f(\mathit{conf}(S, C))
\]

The function $f$ is trained on a dataset of suggestions and their observed outcomes ($Y$) 
so that the resulting $\hat{p}$ provides a more accurate estimate of the true utility. In other words, the objective is for $\hat{p}$ to be as close to a perfectly calibrated estimate as possible, such that $E[Y | \hat{p} = p] \approx p$.

This formal framework allows us to articulate our research goals precisely:
\begin{itemize}
    \item \textbf{RQ1} investigates whether the calibrated probability $\hat{p}$  increased the reliability of confidence signals by having a stronger correlation with the observed developer behavior (our proxy for $Y$) than the raw confidence $\mathit{conf}(S, C)$.
    \item \textbf{RQ2} explores whether the calibration model $f$ becomes more effective when it is specialized, that is, trained on data partitioned by a specific person or project.
    \item \textbf{RQ3} examines how the reliability signal in the form of calibrated probability $\hat{p}$ can be most effectively communicated to developers within the IDE.
\end{itemize}

\begin{figure*}
    \centering
    \includegraphics[width=0.98\linewidth]{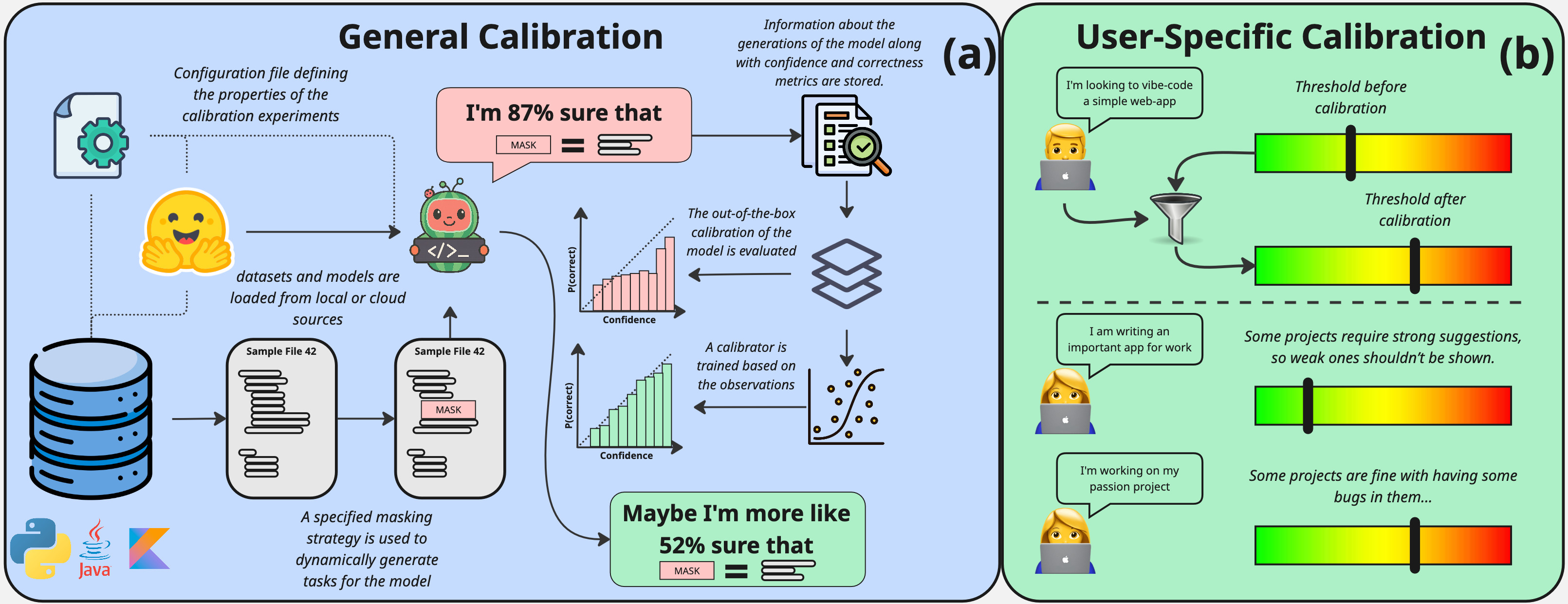}
    \caption{(a) In general calibration, a calibrator is trained on the model's confidence and correctness to get more reliable estimates. (b) In the user-specific stage, the calibrator is adapted to match user- and project-specific needs.}
    \label{fig:calib-pipeline}
\end{figure*}

\section{\textbf{RQ 1.} }
\label{sec:rq1}

In our first research question, we investigate the feasibility of using post-hoc calibration on code completion models. Specifically, we aim to determine (1) if a calibrated confidence score provides a more reliable signal of suggestion utility than the model's raw confidence, and (2) whether specializing the calibration model to a specific usage context 
yields further improvements.

\subsection{Method}
\label{sec:rq1:methods}

To systematically address this question, we developed \textbf{Calibrate-CC}, a scalable and extensible Python framework for evaluating and calibrating code completion models. Our framework, developed with the Hugging Face \textit{transformers} and \textit{datasets} libraries, is intentionally designed to be model- and data-independent. This decision ensures that it can integrate with any open source decoder-based language model and a wide range of code datasets. It supports configurable masking policies, generation parameters, and evaluation strategies, enabling reproducible research into model calibration.

\subsubsection{Data Generation for Calibration}
\label{sec:data_generation}
The first step in our methodology is to generate a dataset suitable for training a calibration model. This process involves creating pairs of model confidence scores and their corresponding ground-truth utility labels:
\[
(\mathit{conf}(S, C), Y)
\]
Our framework automates this pipeline as follows:

\begin{enumerate}
    \item \textbf{Dataset Loading:} We gather code examples from an internal dataset specifically tailored for code completion. This dataset is decontaminated with respect to the model's training data. Although this dataset includes examples from multiple programming languages, our research emphasizes the subsets related to the \textbf{Java}, \textbf{Python}, and \textbf{Kotlin} languages. Each sample in the dataset can be seen as a 4 tuple \[(\mathit{prefix, suffix, middle, multifileContext})\] where the objective is to predict the middle based on the prefix, suffix, and multi-file context\footnote{This is a set consisting of three additional files, each containing code related to the file where the current generation is being developed.}. As per the definition in \autoref{sec:problem}, these fields are combined into a unified piece of context $ctx \in C$ and a suggestion $s \in S$ is generated by the model $M$. \autoref{tabl:datasetdistr} presents details about the data distribution and statistics of the dataset used.

    \begin{table*}[ht]
      \centering
      \begin{tabular}{lccc}
        \toprule
        Language & Total Unique Examples & Average Context Length / Example & Average Masked Span Length \\ 
        \midrule
        Java     & $6151$ & $6788.53 \pm 101.59$ & $64.54 \pm 8.48$ \\
        Python   & $6900$ & $12901.29 \pm192.26 $ & $ 61.02\pm 8.93$ \\
        Kotlin   & $6988$ & $5874.25 \pm 84.76$ & $69.97 \pm 9.12$ \\
        \midrule
        Total    & $ 20039 $ & $ 8574.50 \pm 151.81$ & $ 65.22 \pm 8.87$ \\
        \bottomrule
      \end{tabular}
      \caption{Data distribution and statistics of the dataset used for calibration and evaluation. The statistics presented are based on tokens as ($mean \pm std.$), using the \textit{[MODEL]} tokenizer.}
      \label{tabl:datasetdistr}
    \end{table*}
    
    \item \textbf{Masking and Prompting:} For each code example, a masking policy is applied to create a realistic code completion scenario. For instance, the \textit{RandomLineSpanPolicy} selects a span of lines to be the \textit{ground truth} completion and splits the remaining source code into a prefix ($C_{\mathit{prefix}}$) and a suffix ($C_{\mathit{suffix}}$). This pair forms the context $C$ for a fill-in-the-middle (FIM) prompt.
    
    \item \textbf{Generation and Confidence Extraction:} The context $C$ is passed to the base model $M$ to generate a suggestion $S$. During this process, we record the sequence of token log-probabilities. As formally defined in \autoref{sec:problem}, we calculate the raw model confidence, $\mathit{conf}(S, C)$, as the average log-probability of the generated tokens; equivalent to the geometric mean of the token probabilities in linear space.
    
    \item \textbf{Evaluation and Labeling:} The generated suggestion $S$ is evaluated against the ground truth to produce the outcome variable $Y$. This evaluation constitutes the basis for our ground-truth labels.
\end{enumerate}

\subsubsection{Defining Ground Truth Labels ($Y$)}
When correlating with real-world developer interactions in our main analysis, we define $Y$ as the \textbf{preserved ratio} of the suggestion. This continuous metric, valued in $[0, 1]$, measures the ratio of similarity between the provided completion and 
the code eventually included in the code by the developer. Assume that the expected completion is denoted by $middle$, and the generated completion by $\hat{middle}$. The preserved ratio is calculated as

\[
Y = 1 - \frac{L(middle, \hat{middle})}{max(\mid middle\mid,\mid\hat{middle}\mid)}
\]

where $L(s_1, s_2)$ is the Levenshtein distance metric. In essence, the hypothesis behind opting for this metric over one like \textbf{Exact Match} is that the likelihood of a developer accepting a suggestion correlates directly with how closely the suggestion aligns with the expected completion.

\subsubsection{Training Calibration Models}
With a generated dataset of $(\mathit{conf}(S, C), Y)$ pairs, we can train the calibration model $f$, as formalized in \autoref{sec:problem}. For this study, similar to previous work by Spiess et. al~\cite{spiess2024calibration}, we use Platt Scaling~\cite{platt1999probabilistic}, a parametric method suited for correcting miscalibrations by fitting a logistic regression model. We train two different types of calibrators to address RQ1:

\begin{itemize}
    \item \textbf{General Calibrator ($f_{\mathit{general}}$):} This model is trained on the entire dataset, $\mathcal{D}_{all}$, which aggregates all observations across all programming languages. The goal is to learn a single, universal calibration function that improves confidence estimates regardless of the language context.
    
    \item \textbf{Language-Specific Calibrators ($f_{\mathit{lang}}$):} To test the hypothesis that calibration benefits from specialization, we train separate models for each language. For a given language $L$ (e.g., Python), we first create a filtered dataset $\mathcal{D}_{L} = \{(\mathit{conf}_i, Y_i) \in \mathcal{D}_{all} | \text{language}_i = L\}$. We then train a dedicated calibrator, $f_{L}$, solely on this subset.
\end{itemize}

\subsubsection{Evaluation of Calibration}
To evaluate the performance of our calibration models, we use standard metrics designed to quantify the alignment between the predicted calibrated probabilities ($\hat{p}$) and the observed outcomes ($Y$) on a held-out set of $N$ samples.

\begin{itemize}
    \item \textbf{Expected Calibration Error (ECE):} This metric approximates the difference between expected confidence and accuracy. The predictions are partitioned into $M$ bins ($B_1, \ldots, B_m$) based on their confidence scores. The ECE is the weighted average of the absolute difference between the mean predicted confidence and the mean true outcome in each bin.
    $$ \text{ECE} = \sum_{m=1}^{M} \frac{|B_m|}{N} \left| \left( \frac{1}{|B_m|} \sum_{i \in B_m} Y_i \right) - \left( \frac{1}{|B_m|} \sum_{i \in B_m} \hat{p}_i \right) \right| $$
    A lower ECE indicates better calibration.

    \item \textbf{Brier Score (BS):} The Brier score measures the mean squared error between the predicted probability $\hat{p}_i$ and the outcome $Y_i$. It is a measure of both calibration and resolution.
    $$ \text{BS} = \frac{1}{N} \sum_{i=1}^{N} (\hat{p}_i - Y_i)^2 $$
    Lower values are better. 

    \item \textbf{The Brier Skill Score (BSS):} quantifies the relative improvement in accuracy of one probabilistic forecast over another. 
    Formally, it compares the Brier Scores of two forecasts, $x$ and $y$, as 
    \[
    BSS = 1 - \frac{BS_x}{BS_y},
    \]
    where $BS_x$ and $BS_y$ denote the Brier Scores of forecasts $x$ and $y$, respectively. 
    A positive BSS indicates that forecast $x$ produces lower error (and is therefore more skillful) than forecast $y$, while a negative BSS indicates the opposite. In general:
    \begin{itemize}[leftmargin=0em]
        \item $BSS > 0$: forecast $x$ is more skillful than forecast $y$;
        \item $BSS = 0$: forecast $x$ has no additional skill over forecast $y$;
        \item $BSS < 0$: forecast $x$ is less skillful than forecast $y$.
    \end{itemize}
    In this work, we adopt a \textit{naive base rate model} as the reference forecast $y$, which always predicts the historical average acceptance rate. 
    Accordingly, our reported BSS values represent the relative improvement of each calibrated model over this naive baseline.

    \item \textbf{Maximum Calibration Error (MCE):} Similar to ECE, MCE considers the binned differences but reports the maximum deviation, i.e., the worst-case miscalibration.
    $$ \text{MCE} = \max_{m \in \{1, \ldots, M\}} \left| \left( \frac{1}{|B_m|} \sum_{i \in B_m} Y_i \right) - \left( \frac{1}{|B_m|} \sum_{i \in B_m} \hat{p}_i \right) \right| $$
\end{itemize}

It is important to note that when discussing metrics like the Brier Score in calibration contexts, we are essentially discussing a concept similar to the mean squared error (MSE). In our approach, the positive (1) label indicates when a developer (partially) accepts a suggestion, corresponding to a continuous value from 0 to 1. Our data show that partial acceptance is rare (only $0.059\%$ of the data) therefore we opt to threshold these values (where $x \geq 0.5$ are assigned label 1 and the rest value 0). As such, the data is converted into binary, which justifies using the Brier score in our terminology. We use the average acceptance rate of 26\% for the general model, while for specific languages, we use their distinct average acceptance rates.

\subsection{Results}
\label{sec:rq1:resutls}
To evaluate the efficacy and methodology of our approach, we use a large-scale, real-world dataset of anonymized user interactions. This data subset was collected over a six-month period, from January 1, 2025, to July 1, 2025. The subset contains \textbf{24,767,961} records, obtained from \textbf{759,983 unique devices} across \textbf{3,963,331 distinct sessions}.

This subset of the data contains interaction data across the programming languages of interest. More specifically, Java, Python, and Kotlin. For the purpose of our analysis, we merge interactions from Jupyter notebooks (\textit{jupyterpython}) with standard Python (\textit{python}) files into a single, unified Python category. This decision is motivated by the semantic similarity of the code and the desire to create a comprehensive view of Python development activity. The distribution of records across these languages is presented in Table~\ref{tab:data_distribution}. Java constitutes the largest portion of the dataset (48.20\%), followed by the Python category (38.20\%) and Kotlin (13.60\%).

\begin{table*}[ht]
\centering
\caption{Statistical overview and language distribution of the user interaction dataset.}
\label{tab:data_distribution}
\begin{tabular}{lccc}
\toprule
\textbf{Language} & \textbf{Total Records} & \textbf{Share (\%)} & \textbf{Acceptance Rate (\%)} \\
\midrule
Java & 11,938,231 & 48.20\% & 28.09\%\\
Python* & 9,460,897 & 38.20\% & 24.29\%\\
Kotlin & 3,368,833 & 13.60\% & 23.44\%\\
\midrule
\textbf{Total} & \textbf{24,767,961} & \textbf{100.00\%} & 26.00\%\\
\bottomrule
\multicolumn{3}{l}{\footnotesize{* formed by 8,894,629 python records and 566,268 jupyterpython records.}}
\end{tabular}
\end{table*}

This portion of the data is derived from a selection of suggestion interactions that we can confidently confirm have been presented to the user at least once and have either been dismissed or (partially)\footnote{Partial acceptance involves accepting some tokens from a completion. Unlike tab completion, users can select a token subset as a partial prefix from the suggestion.} accepted by them. In this analysis, we specifically focus on the interaction data with the generation model
\footnote{Mellum-4b Model~\url{https://huggingface.co/JetBrains/Mellum-4b-base}}, 
for which we have complete access to confidence metrics like token-level probability.

\begin{table*}[ht]
\centering
\caption{Calibration results comparing the uncalibrated model confidence, a general calibrator, and language-specific calibrators. The evaluation is performed on both the \textbf{Training Dataset} and a real-world \textbf{Logs Dataset}. Arrows indicate the preferred direction for each metric ($\downarrow$: lower is better, $\uparrow$: higher is better).}
\label{tab:rq1_results}
\begin{tabular}{llcccc|cccc}
\toprule
& & \multicolumn{4}{c}{\textbf{Training Dataset}} & \multicolumn{4}{c}{\textbf{Logs Dataset}} \\
\cmidrule(lr){3-6} \cmidrule(lr){7-10}
\textbf{Language} & \textbf{Model} & \textbf{ECE} $\downarrow$ & \textbf{BS} $\downarrow$ & \textbf{BSS} $\uparrow$ & \textbf{MCE} $\downarrow$ & \textbf{ECE} $\downarrow$ & \textbf{BS} $\downarrow$ & \textbf{BSS} $\uparrow$ & \textbf{MCE} $\downarrow$ \\
\midrule
\multirow{4}{*}{Java} & Uncalibrated & 0.345 & 0.345 & -1.728 & 0.477 & 0.300 & 0.308 & -0.524 & 0.488 \\
 & General ($f_{general}$) & 0.034 & 0.225 & -0.777 & 0.300 & 0.082 & 0.203 & -0.006 & 0.444 \\
 & Language-Specific ($f_{Java}$) & 0.031 & 0.225 & -0.783 & 0.270 & 0.100 & 0.206 & -0.022 & 0.443 \\
 & Baseline (Avg) & - & 0.126 & - & - & - & 0.202 & - & - \\
\midrule
\multirow{4}{*}{Python} & Uncalibrated & 0.408 & 0.392 & -1.644 & 0.487 & 0.298 & 0.288 & -0.567 & 0.491 \\
 & General ($f_{general}$) & 0.070 & 0.230 & -0.551 & 0.489 & 0.096 & 0.187 & -0.018 & 0.469 \\
 & Language-Specific ($f_{Python}$) & 0.053 & 0.228 & -0.536 & 0.500 & 0.083 & 0.185 & -0.005 & 0.478 \\
 & Baseline (Avg) & - & 0.148 & - & - & - & 0.184 & - & - \\
\midrule
\multirow{4}{*}{Kotlin} & Uncalibrated & 0.380 & 0.377 & -2.030 & 0.457 & 0.317 & 0.304 & -0.694 & 0.535 \\
 & General ($f_{general}$) & 0.041 & 0.230 & -0.848 & 0.332 & 0.112 & 0.190 & -0.059 & 0.572 \\
 & Language-Specific ($f_{Kotlin}$) & 0.044 & 0.230 & -0.852 & 0.346 & 0.116 & 0.191 & -0.065 & 0.570 \\
 & Baseline (Avg) & - & 0.124 & - & - & - & 0.179 & - & - \\
\midrule
\multirow[t]{3}{*}{Overall} & Uncalibrated & 0.379 & 0.372 & -1.768 & 0.463 & 0.301 & 0.300 & -0.558 & 0.495 \\
 & General ($f_{general}$) & 0.041 & 0.228 & -0.698 & 0.368 & 0.092 & 0.195 & -0.015 & 0.465 \\
 & Baseline (Avg) & - & 0.134 & - & - & - & 0.192 & - & - \\
\bottomrule
\end{tabular}
\end{table*}

We present a detailed overview of the results from RQ1 in \autoref{tab:rq1_results}. The results are shown for both our internal training dataset and a real-world logs dataset (which in ML terminology would equate to the train and test set). With this data we are able to assess performance on both in-distribution and out-of-distribution data.

\paragraph{Observation 1} The primary observation is that the uncalibrated model consistently exhibits poor calibration as evident by high expected calibration errors (ECE), Brier scores (BS), maximum calibration errors (MCE) and strongly negative Brier Skill Scores (BSS). After applying the general calibrator ($f_{general}$), we see a uniform improvement across all metrics for Java, Python, Kotlin, and the overall aggregation. For example, in the logs data set, the overall ECE drops from 0.301 to 0.092, and the Brier Skill Score increases from –0.558 to –0.015; this indicates that the calibrated predictions perform nearly on par with or slightly
worse than a naive mean probability baseline. This confirms that $f_{general}$ effectively reduces overconfidence and improves probabilistic accuracy, however, is still not
improving on the 
naive baseline.

\paragraph{Observation 2} The second observation concerns the limited benefit of specialization. Comparing the general calibrator ($f_{general}$) to the language-specific variants (e.g., $f_{Java}$, $f_{Python}$, $f_{Kotlin}$) reveals no consistent advantage on the Logs Dataset. While in isolated cases, such as for Python, the language-specific model slightly improves metrics all-round, these marginal gains are accompanied by minor drops in the same metrics for another language-specific model (see BSS $f_{Java}$). This suggests that, although specialization can slightly adapt to language-level idiosyncrasies, the broader statistical exposure of the general model yields more stable calibration performance under distribution shift, which is a likely advantage when facing the heterogeneous and noisy real-world user data.

\observation{\textbf{\textit{Summary of RQ1.}} Calibration improves model confidence reliability across languages and datasets. While the general calibrator ($f_{general}$) consistently outperforms the uncalibrated model, it is on par with or slightly better than a naive baseline. Language-specific variants yield minor, inconsistent gains and highlight the robustness and practicality of a general calibration approach.}

\section{\textbf{RQ 2.} }
\label{sec:rq2}
While RQ1 explores static calibration models, RQ2 investigates the hypothesis that the optimal calibration of a code model is not universal but is instead a function of the individual developer and the specific project context. Developers have unique coding styles, acceptance thresholds (e.g., during \textit{vibe coding} compared to when programming for work), and expertise, which may change depending on the project they are working on. This motivates a shift from static, one-size-fits-all calibrators to dynamic models that adapt over time to these specific contexts.

\subsection{Method}
\label{sec:rq2:methods}
To address this, we introduce an online learning framework where the models are continuously updated based on a stream of developer interactions. This allows us to create and compare several adaptive models, each of which captures a different level of context.

\subsubsection{Online Learning Framework}
In an online learning setting, the data arrives as a sequential stream~\footnote{We define a stream to be a temporally ordered set of AI-completion interaction logs for a given grouping.} of observations. 
At each time step $t$, we observe a new interaction consisting of the model's raw confidence for a given suggestion, $\mathit{conf}_t(S, C)$, and the developer's anonymized feedback, used as the outcome $Y_t$ (e.g, acceptance).

Our calibration model, $f$, evolves over time, denoted as $f_t$. At any given time $t$, the current model $f_t$ produces a calibrated prediction $\hat{p}_t = f_t(\mathit{conf}_t(S, C))$. After the true outcome $Y_t$ is observed, this new pair $(\mathit{conf}_t, Y_t)$ is used to update the model to $f_{t+1}$ for subsequent predictions. We apply this framework to two distinct scenarios to disentangle the effects of user and project context:
\begin{itemize}
    \item \textbf{Per-Person Calibration ($f_{user, t}$):} A single adaptive model is maintained for each developer. This model is updated with all interaction data from that user, \textit{regardless of the project}. This model aims to capture a developer's general, project-agnostic interaction style.
    
    \item \textbf{Per-Person-Per-Project Calibration ($f_{user, project, t}$):} An adaptive model is maintained for each unique \textit{user-project pair}. The model updates only with data from that user's interactions on that project. This tests the hypothesis that a developer's behavior is highly context-dependent (For instance, a user's criteria for Project A are different from those for Project B.).
\end{itemize}

\subsubsection{Progressive Validation and Model Adaptation}
To evaluate the performance of these adaptive models in a temporally-aware manner, we use a progressive validation (or rolling window) methodology. This approach allows us to measure how the calibration performance evolves as the model is exposed to more data. The process is as follows:

\begin{enumerate}
    \item \textbf{Initialization:} Each adaptive model ($f_{user, t}$ or $f_{user, project, t}$) is initialized, for example, with the general-purpose calibrator, $f_{general}$, from RQ1. Let the initial model be $f_0$.

    \item \textbf{Evaluation Window:} We define a sliding window $W$ of a fixed size $k$. At a given point in a data stream represented by time $T$, the window $W_T$ contains the $k$ most recent observations for that stream:
    $$ W_T = \{ (conf_t, Y_t) \}_{t=T-k+1}^{T} $$
    
    \item \textbf{Evaluate:} We use the model as it existed before seeing the data in the current window (i.e., $f_{T-k}$) to make predictions for all $k$ observations within $W_T$. We then compute our standard calibration metrics (ECE, BSS, etc.) over this window. For calculating the Brier Skill Score (BSS), a reference model is required for comparison. Bayesian updates are applied to this reference model. As with RQ1, we use the average acceptance probability as the initial assumption, updating the reference model (average acceptance estimate) with observations from each window. This method ensures a more accurate BSS across windows.
    
    \item \textbf{Fine-tune (Adapt):} The observations within the window $W_T$ are used to update the calibration model from $f_{T-k}$ to a new version, $f_T$. For Platt Scaling, this involves incrementally re-training the logistic regression parameters using a recent history of observations from the relevant stream.
    
    \item \textbf{Slide and Repeat:} The window is moved forward, and the process is repeated. This essentially generates a time series of performance metrics for each adaptive model.
\end{enumerate}

\subsubsection{Temporal Evaluation}
The primary output of our progressive validation is a time series of performance scores for each metric (e.g., $ECE_T$, $BSS_T$). Our analysis of these results will focus on several temporal aspects:
\begin{itemize}
    \item \textbf{Correlation with Behavior:} We measure the overall correlation between the calibrated probabilities ($\hat{p}$) from each type of adaptive model and the outcomes ($Y$), comparing them to each other and to the static models from RQ1.
    
    \item \textbf{Convergence and Stability:} We will analyze the time series of the metric scores to determine if the adaptive models converge to a stable state of improved calibration and how quickly they do so.
        
    \item \textbf{Comparative Performance:} By comparing the average performance of the $f_{user, t}$ and $f_{user, project, t}$ models, we can determine which contextual factors (the person, the project, or the interaction between them) are most important for achieving accurate calibration.
\end{itemize}


\subsection{Results}
\label{sec:rq2:resutls}

Our analysis of the adaptive calibration framework shows that dynamic, user-aware learning offers moderate but consistent gains over the static general-purpose calibrator from RQ1. We evaluated two adaptive strategies: a \textbf{Per-Person} model ($f_{user, t}$) and a more granular \textbf{Per-Person-Per-Project} model ($f_{user, project, t}$).

\autoref{tab:rq2_overall_results} summarizes the aggregated performance across all users and projects. Compared to the general calibrator ($f_{general}$), which achieves an average ECE of 0.092, both adaptive models demonstrate lower calibration errors, with ECE values of \textbf{0.066} and \textbf{0.068}, respectively. Similarly, both models yield reduced Brier Scores and improved Brier Skill Scores (BSS), moving from the negative baseline of –0.015 to positive averages of 0.020 and 0.013. This improvement in BSS, which measures relative performance over a naïve mean-probability predictor, confirms that the adaptive models are not only better calibrated but also more informative in their probabilistic predictions. The comparable performance of the two adaptive strategies suggests that most of the benefit arises from user-level adaptation, with the added project-level granularity contributing marginally at best. This can be attributed to data sparsity: on average, individual users provide longer interaction sequences than user–project pairs, leading to more stable model updates.

\begin{table*}[tbp]
\centering
\caption{Overall performance of adaptive calibration models compared to the static general model. Reported are the mean $\pm$ standard deviation across users/projects. Lower is better for ECE, Brier, and MCE; higher is better for BSS.}
\label{tab:rq2_overall_results}
\begin{tabular}{lcccc}
\toprule
\textbf{Model} & \textbf{Adapted ECE} $\downarrow$ & \textbf{Adapted Brier} $\downarrow$ & \textbf{Adapted BSS} $\uparrow$ & \textbf{Adapted MCE} $\downarrow$  \\
\midrule
General ($f_{\text{general}}$, from RQ1) 
& $0.092$ 
& $0.195$ 
& $-0.015$
& $0.465$ \\
\midrule
Per-Person ($f_{\text{user},t}$) 
& $0.066 \pm 0.038$ 
& $0.166 \pm 0.054$ 
& $0.020 \pm 0.107$ 
& $0.470 \pm 0.253$ \\
Per-Person-Project ($f_{\text{user,project},t}$) 
& $0.068 \pm 0.038$ 
& $0.165 \pm 0.053$ 
& $0.013 \pm 0.128$ 
& $0.484 \pm 0.256$ \\
\bottomrule
\end{tabular}
\end{table*}

To better understand the relationship between activity level and calibration effectiveness, we segmented users and projects into three groups of activity:\textbf{Low}, \textbf{Average}, and \textbf{High}, based on the volume of interactions (\autoref{tab:rq2_segmented_results}). The results reveal a clear dependency between data availability and adaptation quality.

\begin{table*}[ht]
\centering
\caption{Performance of adaptive models segmented by user activity level. Reported are the mean $\pm$ standard deviation across users/projects. Lower is better for ECE, Brier, and MCE; higher is better for BSS.}
\label{tab:rq2_segmented_results}
\begin{tabular}{llcccc}
\toprule
\textbf{Model} & \textbf{Activity Group} 
& \textbf{Adapted ECE} $\downarrow$ 
& \textbf{Adapted Brier} $\downarrow$ 
& \textbf{Adapted BSS} $\uparrow$ 
& \textbf{Adapted MCE} $\downarrow$ \\
\midrule
\multirow{3}{*}{Per-Person} 
 & Low & 
 $0.070 \pm 0.039$ & 
 $0.170 \pm 0.053$ & 
 $0.009 \pm 0.113$ & 
 $0.486 \pm 0.255$ \\
 
 & Average & 
 $0.066 \pm 0.036$ & 
 $0.169 \pm 0.053$ & 
 $0.016 \pm 0.128$ & 
 $0.473 \pm 0.253$ \\

 & High & 
 $0.065 \pm 0.039$ & 
 $0.164 \pm 0.055$ & 
 $0.024 \pm 0.096$ & 
 $0.465 \pm 0.253$ \\
\midrule

\multirow{3}{*}{Per-Project} 
 & Low & 
 $0.095 \pm 0.042$ & 
 $0.179 \pm 0.048$ & 
 $-0.040 \pm 0.201$ & 
 $0.577 \pm 0.263$ \\
 
 & Average & 
 $0.072 \pm 0.036$ & 
 $0.169 \pm 0.052$ & 
 $0.003 \pm 0.138$ & 
 $0.506 \pm 0.257$ \\
 
 & High & 
 $0.065 \pm 0.037$ & 
 $0.163 \pm 0.054$ & 
 $0.012 \pm 0.120$ & 
 $0.473 \pm 0.254$ \\
\bottomrule
\end{tabular}
\end{table*}

For the \textbf{Per-Person} model, ECE and BS steadily improve with increased activity, and the BSS rises from 0.009 for low-activity users to 0.024 for highly active users, indicating that personalization becomes increasingly effective with richer interaction histories. In contrast, the \textbf{Per-Project} model exhibits greater volatility: for low-activity pairs, performance worsens relative to the general model (BSS of –0.040), and reflects unreliable adaptation under data scarcity. However, with sufficient activity, the model recovers and achieves a BSS of 0.012, similar to the per-person model.

\paragraph{Observation 3} We compute the skill score on using model confidence for predicting
code retention (acceptance), over a very large
amount of actual telemetry data. Given the common 
practice of using \emph{model confidence} to select good suggestions, we believe
this is an important quantity to measure; to our knowledge, this hasn't been done
before. Despite Platt scaling, the skill remains rather low, especially when compared to prior reported skill scores,
which were for code \emph{test-passing} correctness~\cite{spiess2024calibration,huang2023look}. 
This finding suggests that using model confidence to select code completions may have
limitations, and other approaches may provide confidence measures that have
more reliable associations with retention.

\observation{\textbf{\textit{Summary of RQ2}.} Our findings for RQ2 demonstrate that adaptive calibration improves both confidence alignment (as captured by ECE and BS) and relative predictive skill (as reflected by BSS). However, these benefits are tightly coupled with the availability of user- or project-specific data. Personalized calibrators become more effective as they observe more interactions, whereas in low-data regimes, the static general model remains the most reliable option.}

\section{\textbf{RQ 3.} }
While RQs 1 and 2 discussed the calibration of model confidence in code suggestions, RQ3 intends to investigate the best way to communicate confidence reliability values to the IDE user.

\label{sec:rq3}

\subsection{Method}
\label{sec:rq3:methods}

To understand how the reliability signal, represented by LLM's confidence, should be integrated into developer workflows, we conducted a multi-phase design user study. It combined scenario-based design~\cite{rosson2007scenario} with semi-structured interviews~\cite{blandford2013semi} and larger-scale surveys to systematically explore and validate implementation approaches for confidence signals in AI-assisted code generation. The study received ethical approval from [Institution] Human Research Ethics Committee (application number: 5682).\footnote{Institution name has been anonymized for double-blind review and will be included in the camera-ready version.}

\subsubsection{Design Sessions with Professional Designers}
We conducted 90-minute design sessions with three expert UI designers (14, 7, and 15 years of experience) specializing in AI interfaces (1, 1.5, and 3 years of experience correspondingly). Participants were recruited from a technology [Company]\footnote{Company name anonymized for review.} where they were employed as professional designers. The sessions employed collaborative sketching~\cite{sanders2008co} to explore interface concepts for presenting confidence information. The protocol of sessions was reviewed by professional copyeditors employed by the [Company] and refined through pilot testing. The protocol included system introduction, structured design tasks, and reflection on the suggested design of confidence signal integration into the UI of the in-IDE AI assistant. Sessions were conducted remotely, and audio and video were recorded for analysis. The first two authors collaboratively reviewed session transcripts and design artifacts compiled in Figma\footnote{Figma: Collaborative Design Tool \url{https://www.figma.com}} to identify actionable design principles and interface concepts. Through iterative discussion, the researchers consolidated outputs into design dimensions of interface approaches to be further explored with professional developers.

\subsubsection{Semi-structured Interviews with Developers}
\begin{figure}[tbp]
  \centering
  \begin{subfigure}{\linewidth}
    \centering
    \includegraphics[width=\linewidth]{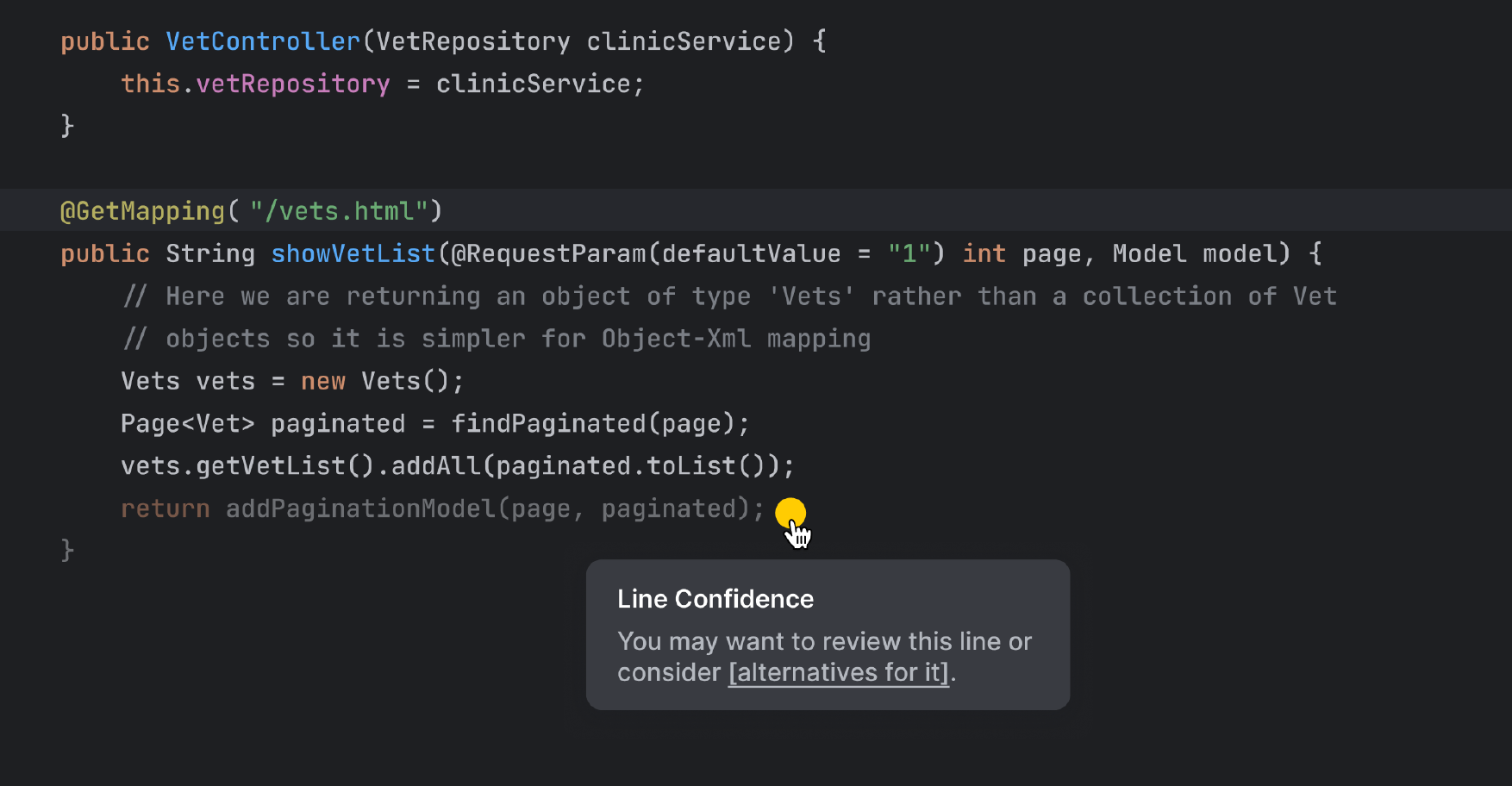}
    \caption{In-editor autocompletion with reliability signal}
    \label{fig:autocompl}
  \end{subfigure}
  
  \vspace{0.5em}
  
  \begin{subfigure}{\linewidth}
    \centering
    \includegraphics[width=\linewidth]{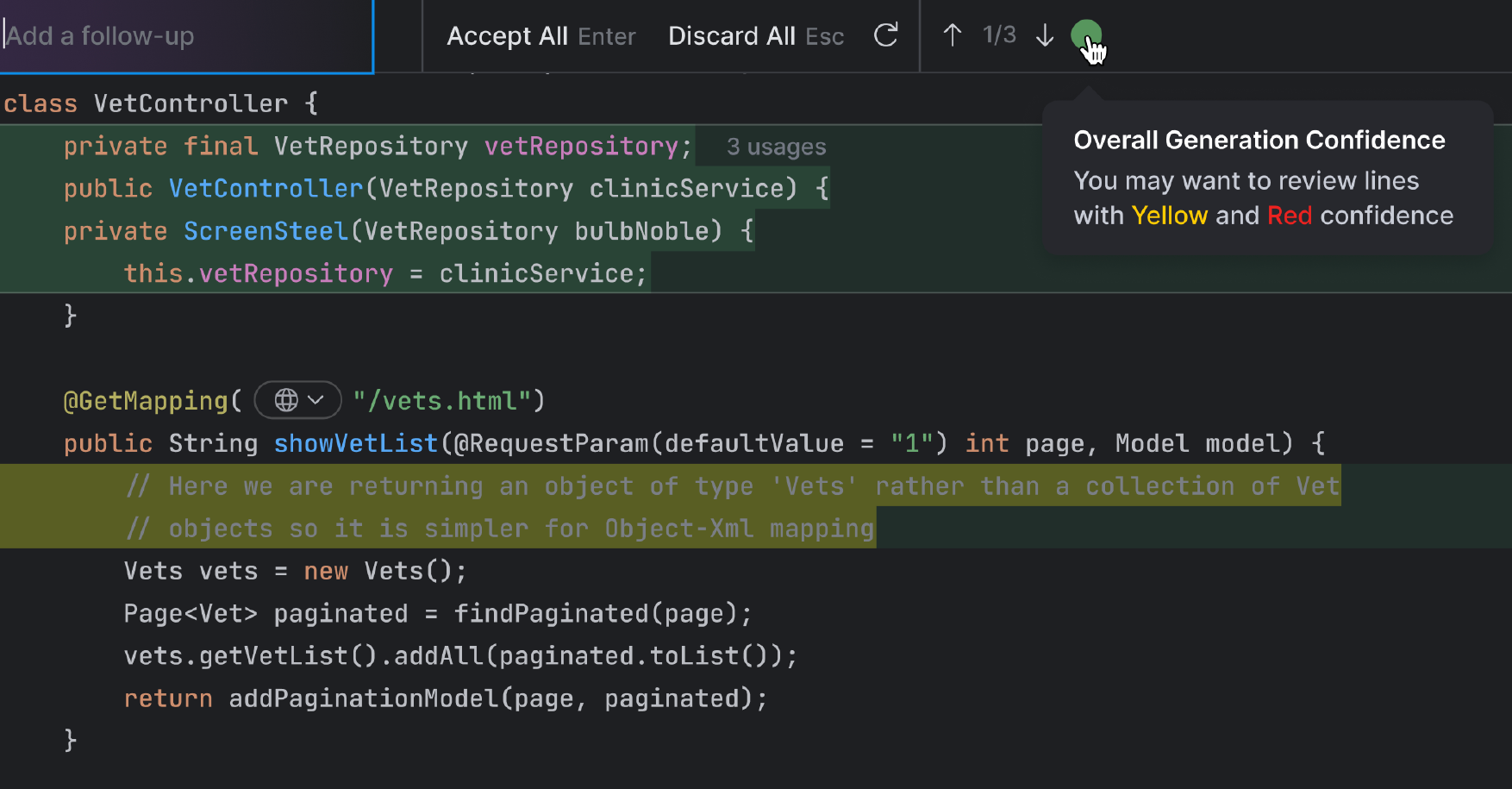}
    \caption{In-editor code generation with reliability signal}
    \label{fig:ineditor}
  \end{subfigure}
  
  \caption{Candidate interface designs for reliability signal presentation. (a) A tooltip appears when hovering over the yellow signal, prompting review. (b) Yellow highlighting indicates medium-reliability lines in the generation panel.}
  \label{fig:signal-ui}
\end{figure}

Five professional developers (3-6 years of experience, using AI coding tools daily) were recruited via internal advertisement within a [Company] and participated in 30-minute semi-structured interviews. The interviews explored preferences regarding design dimensions identified in the previous phase. Participants were introduced to the calibrated confidence concept as a reliability signal and asked to identify contexts where they would be useful, appropriate granularity levels for assessment, and potential challenges associated with such signals. Interview recordings were transcribed and collaboratively analyzed by the first two authors to identify preference patterns and integration requirements. Based on this analysis, two candidate interface designs were developed (see~\autoref{fig:signal-ui}) for quantitative validation through a survey.

\subsubsection{Survey Validation}
An online survey was conducted to validate the interface designs with 153 professional developers. Participants were recruited from an opt-in research participant list maintained by the [Company]. Participants received an opportunity to enter a prize draw for one of five 50 USD Amazon eGift Cards or equivalent-value [Company] product as compensation. Participants evaluated both interface approaches (autocompletion-integrated and code generation-integrated) using 5-point Likert scales measuring perceived usefulness, trust impact, and workflow disruption potential, and indicated overall preference. The survey included screening questions to confirm professional coding experience and current use of AI coding tools at least once per month. Quantitative responses were analyzed using descriptive statistics. Open-ended responses provided additional context for interpreting quantitative findings.

\subsection{Results}
\label{sec:rq3:resutls}
The designer phase formulated design considerations for in-IDE AI-generated code reliability signals. Designers suggested that signals could be shown at various workflow points with AI: in-editor autocompletion, in-editor code generation from in-line prompts, and AI chat. Designers also proposed that signals could be attributed to different granularity levels: token, line, several lines, or whole code snippet. Multiple signal variations were suggested: traffic-light-like, numerical, and highlight approaches. Additional considerations included what reliability levels are useful to signal and what explanations might be beneficial.

Overall principles for such a feature, advised by participating designers, are:
\begin{itemize}
\item Leverage Existing Patterns: Build on established IDE conventions and visual signals
\item Minimalism First: Avoid cognitive overload and visual clutter, show minimal information by default with detailed information on demand
\item Actionability: Reliability signals should guide users toward next steps
\item Contextual Adaptation: Different presentations for different user types, code criticality, and granularity levels
\end{itemize}

With distilled questions and design principles established, professional developers showed during interviews a preference for in-editor reliability signals with multi-line granularity. Developers indicated that scores alone provide insufficient actionable information for code acceptance decisions. Participants consistently favored explanatory feedback over numeric confidence indicators and requested reasoning behind code suggestions, echoing the actionability principle from designers. All participants emphasized the critical importance of user-configurable thresholds.

Based on these preferences, we constructed two UI prototypes reflecting the most promising approaches from our design exploration: an autocompletion-integrated signal and an in-editor generation signal (see ~\autoref{fig:signal-ui}). These prototypes were presented to survey participants for quantitative validation.

The survey results revealed that both interface approaches received identical median ratings across key metrics: usefulness (3/5), trust impact (3/5), and workflow disruption potential (2/5). This convergence suggests that neither approach presents significant usability barriers, yet neither emerged as clearly superior in isolation.

However, when forced to choose between approaches, developers demonstrated a clear preference hierarchy. The in-editor generation approach garnered support from 76 participants (51.4\%), while the autocompletion-integrated approach attracted 55 participants (37.2\%). Notably, 17 participants (11.5\%) rejected both options, indicating that reliability signaling may not universally appeal to all developers. This preference for in-editor generation aligns with our interview findings, where developers emphasized the need for actionable information during active code creation. This approach allows developers to assess reliability while maintaining creative control over their code, supporting the design principles identified in our design phase.

\observation{\textbf{\textit{Summary of RQ3.}} Our findings support implementing reliability signals through non-numerical, color-coded indicators integrated into the in-editor generation workflow, as illustrated in ~\autoref{fig:signal-ui} (b). This approach balances visibility with minimal cognitive overhead while providing contextual information at the point of decision-making.}

\section{Discussion}
\label{sec:discussion}

Our mixed-methods approach, combining large-scale telemetry analysis with a user-centered design study, provides a holistic view of confidence calibration for AI coding assistants. Our work provides the first end-to-end treatment of confidence calibration for production-level AI coding assistants, bridging technical calibration methods with human-centered interface design. While prior work has either focused on uncertainty communication strategies~\cite{vasconcelos2025generation} or calibration techniques in isolation~\cite{spiess2024calibration}, we demonstrate how these pieces fit together in a production system. We further discuss the implications of our findings, acknowledge the limitations of our study, and outline threats to its validity.

\subsection{Implications of Findings}
\textbf{General Calibrator Has Low Predictive Ability (RQ1).}
Results in \autoref{tab:rq1_results} demonstrate that the general calibrator ($f_{general}$) consistently improves calibration metrics (such as ECE) over the uncalibrated model across all languages. Notably, it matches or even outperforms language-specific models ($f_{lang}$) in stability when tested on the logs dataset. For platform providers, this is a significant finding as it simplifies deployment and maintenance as one robust calibrator can be used for all users. This study supports \citet{spiess2024calibration}'s findings on the necessity of post hoc calibration by using extensive real developer interaction data instead of synthetic benchmarks. It also indicates that the statistical traits of miscalibration likely apply across languages.

Our findings also highlight a key limitation: calibration aligns model confidence with outcomes, reducing ECE, but fails to make confidence a reliable predictor of user behavior. As indicated in \autoref{tab:rq1_results}, the general calibrator's Brier Skill Score is -0.015, showing that it offers no predictive benefit over a naive average acceptance rate baseline. This suggests a disconnect: model confidence (average token probability) may indicate syntactic or semantic correctness, but does not align with what developers \textit{value} or \textit{accept}. Relying solely on model confidence, even when calibrated, is inadequate for selecting useful code completions.

\textbf{Calibrator Personalization Requires Substantial Data (RQ2).}
While the static general model lacks predictive skill, our results for RQ2 indicate that \textbf{adaptive, personalized calibration can succeed.} According to \autoref{tab:rq2_overall_results}, the per-person model ($f_{\text{user},t}$) achieves an average BSS of 0.020, improving predictive skill over both the static model (BSS -0.015) and the naive baseline. Adapting to individual user acceptance patterns improves the confidence score's significance. However, our findings highlight the challenges in implementing personalization. While the \textit{cold start} problem is evident, it is complex. \textbf{User-level adaptation ($f_{\text{user},t}$) is effective.} Even the "Low" activity group in \autoref{tab:rq2_segmented_results} sees improvement with a BSS of 0.009 over the general model. However, \textbf{project-level adaptation ($f_{\text{user,project},t}$) is risky.} For Low activity user-project pairs, BSS drops to -0.040, indicating that too specific a model without adequate data worsens user experience.

Findings suggest a hybrid approach: initially, all users use the system without any type of calibration.
Once enough interaction data is gathered, the system can transition to a \textbf{per-person model} ($f_{\text{user},t}$) for improved predictions. The detailed \textit{per-person-per-project} model is more viable for users with high interaction within a specific project, where its performance (BSS 0.012) is better.

\textbf{Reliability Signal as a Supplementary Feature (RQ3).} Perhaps our most actionable finding is that the value of a perfectly calibrated model is lost if its reliability is not communicated effectively. Developers are not statisticians, they do not want to interpret raw probability scores. The strong preference for non-numerical, color-coded, and workflow-integrated comprehensive signals provides a clear design guideline: reliability indicators should be informative and intuitive. Furthermore, the relatively modest median scores across all metrics according to our survey suggest that reliability signaling represents a \textit{supplementary feature}. Developers view such signals as potentially useful but not essential, consistent with their expressed preference for user-configurable systems that can be adapted to individual workflows and code criticality levels. 

\subsection{Threats to Validity}

\textbf{Internal Validity.}
Our primary measure of suggestion utility for the log analysis was the \textit{preserved ratio}. While this is a common proxy for acceptance, it is imperfect. A developer might discard a correct suggestion because they thought of a better or alternative implementation. Conversely, they might accept an incorrect suggestion, only to fix it later. This potential mismatch between our proxy and the ground truth is a threat to internal validity.

In RQ3, selection bias may affect our findings, as design session and further validation participants were recruited from a single technology company, and survey participants from an opt-in research pool. However, we deliberately employed multi-phase design to reduce dependence on any single participant pool.

\textbf{External Validity.}
Our study was conducted using a single, specific large language model 
with 4b parameters and on a user base from a single ecosystem\footnote{both of which have been anonymized for the sake of the review.}.
While large and diverse, these results may not generalize perfectly to other models (e.g., GPT-4, Claude 3) which may have different inherent calibration properties, or to developers who exclusively use other editors
and have different workflow habits. Furthermore, our analysis focused on Java, Python, and Kotlin; the findings may not extend to languages with vastly different paradigms, such as C++ or Haskell. The design study participants (RQ3) represented experienced professionals. While findings may not generalize to novice developers or hobbyist programmers, our participant pool aligns with the primary user base for AI coding assistants in professional settings.

\textbf{Construct Validity.}
The survey employed static mockups (Figure~\ref{fig:signal-ui}) to represent proposed interfaces. Static representations cannot fully capture the dynamic interplay between reliability signals and actual coding activities, as participants evaluated designs based on imagined workflows rather than lived experience. However, this approach is standard practice in formative design research~\cite{lim2008anatomy} and appropriate for early-stage exploration before committing resources to full implementation. Our goal was not to validate final designs but to identify promising directions and eliminate clearly problematic approaches. Future work should include field deployment with behavioral logging to validate whether stated preferences translate to actual usage patterns and whether shown reliability signals improve code acceptance decisions in practice.

\section{Conclusion}
\label{sec:conclusion}
The integration of LLMs into IDEs has introduced a new paradigm for software development, but their fallibility raises challenges for reliability and
collaboration. A model's ability to signal when it is uncertain is as important as its ability to generate correct code.

In this paper, we addressed this challenge from both a technical and a human-computer interaction perspective. Through a large-scale analysis of over 24 million real-world interactions, we observe that although post-hoc calibration through Platt-scaling improves the reliability of confidence over the base model, it falls short in predictive power. We found that general
models have low skill, and that while personalization can offer further gains, its benefits are conditional on having a high volume of user data. Finally, our user study with 153 professional developers provided design directions for communicating these reliability signals, showing a preference for intuitive, non-numerical indicators embedded within the primary code generation workflow.

Taken together, our results indicate the need to explore other technical approaches (e.g., use of other internal model signals) to develop AI coding assistants that are more transparent, trustworthy, and user-friendly. Furthermore, future work should validate the provided UI designs in live A/B testing environments and explore how reliable confidence signals impact downstream developer behaviors, such as code review effort and bug introduction rates.
\bibliographystyle{ACM-Reference-Format}
\bibliography{refs}

\end{document}